\documentclass[]{JHEP3}
\usepackage{epsfig}
\usepackage{graphicx}
\def\be{\begin{equation}}
\def\ee{\end{equation}}
\def\bea{\begin{array}}
\def\eea{\end{array}}
\def\beqa{\begin{eqnarray}}
\def\eeqa{\end{eqnarray}}
\def\beqas{\begin{eqnarray*}}
\def\eeqas{\end{eqnarray*}}

\def\bp{\begin{picture}}
\def\ep{\end{picture}}
\def\bc{\begin{center}}
\def\ec{\end{center}}
\def\bfig{\begin{figure}}
\def\efig{\end{figure}}

\def\bit{\begin{itemize}}
\def\eit{\end{itemize}}
\def\nn{\nonumber}
\def\f{\frac}

\def\[{\left[}
\def\]{\right]}
\def\({\left(}
\def\){\right)}

\def\..{\left.}
\def\.{\right.}
\def\tl{\tilde}
\def\ra{\rightarrow}
\def\la{\leftarrow}

\def\NPB#1,{{ Nucl.\ Phys.\ B }{\bf #1},}
\def\PLB#1,{{ Phys.\ Lett.\ B }{\bf #1},}
\def\EPJC#1,{{ Eur.\ Phys.\ Jour.\ C }{\bf #1},}
\def\PRD#1,{{ Phys.\ Rev.\ D }{\bf #1},}
\def\PRL#1,{{ Phys.\ Rev.\ Lett.\ }{\bf #1},}
\def\MPLA#1,{{Mod.\ Phys.\ Lett.\ A }{\bf #1},}

\def\da{\dagger}

\def\la{\lambda}

\def\al{\alpha}

\def\ka{\kappa}

\def\ep{\epsilon}

\def\pr{\prime}

\title{Split Supersymmetry Under GUT and Current Dark Matter Constraints}

\author{Fei Wang$^{1,3}$, Wenyu Wang$^{2,3}$, Jin Min Yang$^{3}$\\
$^1$ Department of Physics and Engineering, Zhengzhou University, 450000,ZhengZhou
P.R.China\\
$^2$ Institute of Theoretical Physics, College of Applied Science,
Beijing University of Technology, Beijing 100124, China\\
$^3$ Institute of Theoretical Physics China, Chinese Academy of Sciences, Beijing, 100080
P.R.China.
}

\abstract{We recalculate the two-loop beta functions for three gauge couplings taking into account all low energy threshold corrections
in split supersymmetry (split-SUSY) which assumes a very high scalar mass scale $M_S$.
 We find that, in split-SUSY with gaugino mass unification assumption
and a large $M_S$, the gauge coupling unification requires a lower bound on gaugino mass.
Combined with the constraints from the dark matter relic density and direct detection limits,
we find that split-SUSY is very restricted and for dark matter mass below 1 TeV
the allowed parameter space can be fully covered by XENON-1T(2017).
}

\begin{document}
\maketitle \indent
\newpage
\section{Introduction}
   It is well known that both the ATLAS and CMS collaborations
have established the existence of a 125 GeV Standard Model (SM)-like Higgs
boson \cite{atlas,cms}.
So far the LHC Higgs data (with large uncertainties) agree well with the SM predictions.
Still, such a newly discovered Higgs boson (especially its enhanced diphoton signal rate
reported by ATLAS) has been interpreted in various new physics frameworks,
among which a particular interesting scenario is low energy supersymmetry \cite{diphoton-susy}.

Supersymmetry (SUSY) is interesting in many aspects.
A very interesting observation is that the observed Higgs boson
mass of 125 GeV falls within the narrow window  $115-135$ GeV
predicted by the Minimal Supersymmetric Standard Model (MSSM).
Besides, the unification of gauge couplings \cite{gutmssm,su5},
which cannot be achieved in the SM,
can be successfully realized by introducing supersymmetric particles with proper
quantum numbers.
The observed cosmic dark matter, which has no interpretation in the SM,
can be perfectly explained in SUSY.

Although SUSY is appealing, no signals of SUSY have been found at the LHC, which
implies that squarks and gluinos should beyond the 1 TeV range.
In fact, the LHC data set a limit\cite{cmssm1,cmssm2} $m_{\tilde g} > 1.5$ TeV
for $m_{\tl{q}} \sim m_{\tl{g}}$
and $m_{\tl{g}}\gtrsim 1$ TeV for $m_{\tl{q}} \gg m_{\tl{g}}$ within the popular CMSSM model.
On the other hand, radiative electroweak symmetry breaking conditions to give a 125 GeV Higgs requires an electroweak fine-tuning (EWFT).
Such a fine-tuning may indicate that we should not expect SUSY to provide naturalness.
Actually, from the viewpoint of quantum field theory, the naturalness problem of the Higgs mass
appears to be quite similar to the cosmological constant problem, since both of them are
related to ultraviolet power divergences. Maybe we can apply the naturalness criterion
of the cosmological constant to SUSY. Split supersymmetry (split-SUSY), proposed
in \cite{nima,giudice1,giudicenima}, gives up naturalness while keeps the other two main virtues:
the gauge coupling unification and viable dark matter candidates.
This split-SUSY scenario assumes a very high scalar mass scale $M_S$ and at low energy
the supersymmetric particles are only the gauginos and higgsinos as well as a
fine-tuned Higgs boson.
With very heavy sfermions this scenario can obviously avoid the flavor problem.

Given the significant progress of the LHC experiment and dark matter
detections \cite{planck,wmap,xenon}, we in this work check the dark matter
and gauge coupling unification in split-SUSY.
In fact, as shown in \cite{cwy,ruderman,nicolas}, the previous dark matter data can already set
some constraints on the parameter space of split-SUSY. The gauge coupling unification
in split-SUSY had been checked at two loop level in a special case assuming $M_1=M_2=M_3=\mu$\cite{giudice1,bernal} and also in complete two loop level in \cite{giudice2}.
We recalculate the two-loop beta functions for three gauge couplings at two loop level taking into account all threshold corrections to check the status of split SUSY after higgs discovery, in particular the gauge coupling unification constraints on dark matter phenomenology.

This paper is organized as follows.  In Sec. \ref{gut} we study the gauge coupling unification
in split-SUSY.
In Sec. \ref{dm} we examine the constraints of dark matter relic density and
direct detections on split-SUSY. Sec. \ref{conclusion} contains our conclusions.

\section{Constraints of Split SUSY From Gauge Coupling Unification}
\label{gut}
 We firstly brief review the split supersymmetry scenario and explain our conventions. More details can be found in \cite{nima,giudice1}.
 The Lagrangian of Split supersymmetry is given by
\beqa
{\cal L}&=&m^2 H^\dagger H-\frac{\lambda}{2}\left( H^\dagger H\right)^2
-\left[ h^u_{ij} {\bar q}_j u_i\epsilon H^*
+h^d_{ij} {\bar q}_j d_iH
+h^e_{ij} {\bar \ell}_j e_iH \right. \nonumber \\
&&+\frac{M_3}{2} {\tilde g}^A {\tilde g}^A
+\frac{M_2}{2} {\tilde W}^a {\tilde W}^a
+\frac{M_1}{2} {\tilde B} {\tilde B}
+\mu {\tilde H}_u^T\epsilon {\tilde H}_d \nonumber \\
&&\left. +\frac{H^\dagger}{\sqrt{2}}\left( \tl{g}_u \sigma^a {\tilde W}^a
+\tl{g}_u^\pr {\tilde B} \right) {\tilde H}_u
+\frac{H^T\epsilon}{\sqrt{2}}\left(
-\tl{g}_d \sigma^a {\tilde W}^a
+\tl{g}_d^\pr {\tilde B} \right) {\tilde H}_d +{\rm h.c.}\right] ,
\label{lagr}
\eeqa
with $\epsilon=i\sigma_2$ and the higgsino components
$\tl{H}_{u,d}$, the gluino $\tl{g}$, the Wino $\tl{W}$, the Bino $\tl{B}$ as well as all the standard model particles with one Higgs doublet $H$. The standard model higgs doublet is the linear combination of two higgs doublets $H=-\cos\beta \epsilon H_d^*+\sin\beta H_u$ which are fine-tuned to have small mass. The definition of scalar quartic coupling $\la$ and the yukawa couplings $h^{u,d,e}_{ij}$ will be given shortly. The parameter $\mu$ arises from the $\mu$-term of the supersymmetric standard model and acts as the higgsino mass parameter.

 The squarks, sleptons, charged as well as the pseudoscalar Higgs from the supersymmetric standard model in split SUSY scenario  are
assumed to be heavy (so that they will not cause a problem in SUSY flavor problems etc) and their masses are assumed to be degenerated at mass scale $M_S$. The coupling constants appeared in previous Lagrangian at the scale $M_S$ are obtained by matching them with the interaction terms of the supersymmetric Higgs doublets $H_u$ and $H_d$
\beqa
{\cal L}_{\rm susy}&=&
-\frac{g^2}{8}\left( H_u^\dagger \sigma^a H_u + H_d^\dagger \sigma^a H_d
\right)^2
-\frac{g^{\prime 2}}{8}\left( H_u^\dagger H_u - H_d^\dagger  H_d
\right)^2 \nonumber \\
&&+\lambda^u_{ij}H_u^T\epsilon {\bar u}_i q_j
-\lambda^d_{ij}H_d^T\epsilon {\bar d}_i q_j
-\lambda^e_{ij}H_e^T\epsilon {\bar e}_i \ell_j
\nonumber \\
&&-\frac{H_u^\dagger}{\sqrt{2}}\left( g \sigma^a {\tilde W}^a
+g^\prime {\tilde B} \right) {\tilde H}_u
-\frac{H_d^\dagger}{\sqrt{2}}\left(
g \sigma^a {\tilde W}^a
-g^\prime {\tilde B} \right) {\tilde H}_d +{\rm h.c.}~.
\label{lagrs}
\eeqa
Because one Higgs doublet can be fine-tuned to be small, the new coupling constants at the scale $M_S$ can be obtained by
replacing  $H_u\to \sin\beta H$ and $H_d\to \cos\beta \epsilon H^*$ into (\ref{lagrs}) with:
\beqa
 \lambda(M_S )&=& \frac{\left[ g^2(M_S )+g^{\prime 2}(M_S )
\right]}{4} \cos^22\beta ,
\label{condh}\\
h^u_{ij}(M_S )&=&\lambda^{u*}_{ij}(M_S )\sin\beta ,
h^{d,e}_{ij}( M_S )=\lambda^{d,e*}_{ij}( M_S )\cos\beta ,\\
\tl{g}_u ( M_S )&=& g ( M_S )\sin\beta ,
\tl{g}_d (M_S )= g (M_S)\cos\beta ,\\
\tl{g}_u^\pr ( M_S )&=& g^\prime ( M_S ) \sin\beta ,
\tl{g}_u^\pr ( M_S )= g^\prime ( M_S )\cos\beta .
\label{condg}
\eeqa
We should note that such tree level relation will hold in higher order only if $\overline{DR}$ (Dimensional Reduction) renormalization scheme is used. Supersymmetry ensures that the gaugino coupling $\hat{g}$ within $\sqrt{2}\hat{g}\phi^i(t^A)_i^j(\psi_j\la^A)$ is equal to the gauge couping $g$. Due to the fact that $\overline{MS}$ is not supersymmetry preserving, the relation $\hat{g}=g$ is spoiled in this scheme.  The relation (\ref{condh}) will be modified \cite{martin} to act as the input of RGE running (see appendix).

Let us take a look at the free parameters in split-SUSY.
It is well known that for the ratios of gaugino masses and gauge couplings
we have
  \beqa
  \f{d}{d\ln\mu}\(\f{M_i}{g_i^2}\)=0
  \eeqa
and thus the ratios are RGE-invariant (up to one-loop level).
This leads to a mass relation given by
 \beqa
 \f{M_1}{g_1^2}=\f{M_2}{g_2^2}=\f{M_3}{g_3^2}=\f{M_U}{g_U^2}, \label{gaugino}
 \eeqa
 with universal gaugino mass at the GUT scale.
 This gaugino mass relation can naturally appear in the ordinary SUSY-SU(5) GUT
models (it can be spoiled by the
introduction of certain higher dimensional representation Higgs fields, e.g.,
the {\bf 75, 200} dimensional Higgs fields \cite{nugaugino1,nugaugino2}).
The two-loop corrections to the mass ratios $M_i/g_i^2$ are subdominant and make negligible contributions to two-loop RGE running of gauge couplings.
So in our following analysis we adopt this gaugino mass relation.
With this mass relation, the low energy SUSY mass
parameters in split-SUSY can be reduced to: $M_3$, $\mu$ and $M_S$. The parameter $\tan\beta$ is chosen by random scan so as to give the 125 GeV higgs in the next section. It was chosen as a free parameter in this section.
To avoid the SUSY flavor problem, split-SUSY assumes $M_S\gg (M_3,\mu)$
and the value of $M_S$ is typically chosen to be higher than 100 TeV.  We should note that the gaugino mass relation will no longer be valid below $M_S$ due to the split nature of the split supersymmetry spectrum. However, various constraints, especially the 125 GeV higgs discovery by LHC, exclude the high $M_S$ scenario and favor scalar superpartners in the region $M_S\sim 10^4-10^8${GeV}\cite{giudice2}. So it can be reasonable to keep the approximate ratio of the gaugino mass relations.

 Preserving gauge coupling unification is one of the two motivations of split-SUSY
which, on the other side, is a highly non-trivial constraint on split-SUSY.
In general, the successful gauge coupling unification at one-loop level taking into
account threshold corrections disfavors a large $M_S$ due to the prediction of
a relatively lower $\al_s(M_Z)$ than the experimental value.
In \cite{nima} it is argued that
the two-loop renormalization group equation (RGE) running
can alleviate this difficulty by pushing up the predicted $\alpha_3(M_Z)$
to around $0.130$ and thus can push up $M_S$ to a large value.
So the inclusion of two-loop RGE runnings for gauge couplings are necessary in order to
achieve the gauge coupling unification in split-SUSY.

In this work we use the method in \cite{twoloop1,twoloop2} to
calculate the two-loop beta functions for three gauge couplings
in split-SUSY, taking into account the threshold corrections.
The results of \cite{giudice1}, which assuming $M_1=M_2=M_3=\mu$,
is a special case of our general results (we checked that in this special
case both results are in agreement).
To study the RGE running for gauge couplings, we also calculated the one-loop beta functions for
Yukawa couplings and gaugino couplings with threshold corrections. There are in total four different scenarios
depending on the relative size of the gaugino masses and $\mu$. The full analytic expression for the beta function in these scenarios can be seen in the appendix. Although the proton decay problem in the split susy scenario will ameliorated, natural doublet-triplet(D-T) splitting may still need certain mechanism.
Incorporating various D-T splitting mechanism can lead to uncertainties in the GUT theory field contents and consequently new matter threshold uncertainties.
So in our study on gauge coupling unification, we neglect possible GUT scale threshold corrections and possible new gauge kinetic terms from  Planck-scale suppressed non-renormalizable operators involving various high representation higgs fields of GUT gauge group. It is well known that the two loop RGE running for gauge couplings are scheme independent, so we use the $\overline{MS}$ couplings in our studying of the gauge coupling unification.

  With the two-loop RGE running of gauge couplings, we can study the gauge coupling
unification requirement for the three free mass parameters in split-SUSY.
To make our calculation reliable, the GUT scale must be significantly
lower than the Planck scale so that the gravitational effects can be neglected.
On the other hand, the GUT scale can not be very low; otherwise it will lead to
fast proton decay.

Note that in ordinary SUSY-GUT, the dominant proton decay comes
from the dimension-5 operators involving the triplet Higgs
and gaugino loops (these dimension-5 operators induce the decay
$p\ra K^+ +\bar{\nu}$, whose experimental bound is
$\tau_{p\ra K^+\bar{\nu}}>3.3\times 10^{33}$ years\cite{protondecay,protondecay2}).
 Since this decay also involves sfermions in the loops,
it is much suppressed in split-SUSY due to very heavy sfermions.
In fact, as noted in \cite{giudicenima}, the contribution from the model-dependent dimension-5 operator which is suppressed by $M_S^4$
is subdominant to dimension-6 operators if the amplitude is suppressed by two light quark/lepton masses.
 In Split Supersymmetry, the heavy squarks can provide adequate suppression and the suppression of
light fermion masses can even be unnecessary.

So for proton decay, we only consider the decay mode $p\ra e^++\pi^0$
induced by the heavy X, Y gauge bosons of SU(5) with mass $M_{GUT}$
through the dimension-6 operators (via gauge boson exchange)\cite{giudice1}:
\beqa
\label{gutlife}
\tau (p\to \pi^0 e^+)=  \left(\frac{ M_{GUT} } {10^{16}~{\rm GeV} } \right)^4  \left( \frac{1/35}{\alpha_{GUT}} \right)^2
\left(\frac{0.015~{\rm GeV}^3}{\alpha_N}\right)^2\left(\frac{5}{A_L}\right)^2~4.4\times 10^{34}~{\rm yr}.\nn
\eeqa
with $A_L$ the operator renormalization factors and $\alpha_N$ the hadronic
matix element.  The lattice result\cite{protondecaylattice} gives $\alpha_N=0.015~{\rm GeV}^3$.

 Combining with the experimental bound  given by\cite{protondecay,protondecay2}
  \beqa
  \tau(p\ra e^++\pi^0)>1.0\times 10^{34} {\rm years},
  \eeqa
 we can find the lower limit for the GUT scale. Taking into account the upper limit (Planck scale) and choosing the central value of $A_L=5$ in equation ($\ref{gutlife}$), the GUT scale should lie in the range
\small
\beqa
\label{guttime}
1.0\times 10^{19}{\rm GeV} >M_{GUT}
>\sqrt{35\al_{GUT}}\( 6.9\times 10^{15} \){\rm GeV}~.
\eeqa
\normalsize
In our numerical study, we require that successful grand unification should satisfy this constraint on the GUT scale.

  The following setting is used in our numerical studies:
  We use the central value of $g_1,g_2$ and $3\sigma$ range of $g_3$ as the input at the electroweak scale.
Other couplings at the electroweak scale, for example, the top yukawa $h_t$ etc,  are extracted from the standard model inputs taking into account the threshold corrections. Relevant details can be seen in the appendix.
We also use their central values in our numerical studies.

  Gauge couplings unification requires that the three gauge couplings meet at the same point with $g_1(M_{GUT})=g_2(M_{GUT})=g_3(M_{GUT})$ and the GUT scale satisfied the equation (\ref{guttime}).
However, in numerical studies, it is not possible to obtain exact equality which differs dramatically from the approach of the one-loop case.
 Because of the decoupled nature of the one-loop gauge couplings running, the unification scale is determined by the intersection of $g_1,g_2$ and one can extrapolate back to predict
$g_3$ at the elctroweak scale. In case of the two loop results, the two-loop RGE running of gauge couplings which amount to numerically solve a series of coupled differential equations are
 obtained from the values at electroweak scale and evolve step by step to GUT scale.
 We thus use the criteria that the gauge couplings unification is satisfied when the three couplings differ within the range 0.005 (less than 1\% error).

\begin{figure}[htbp]
\begin{center}
\includegraphics[width=3.5in]{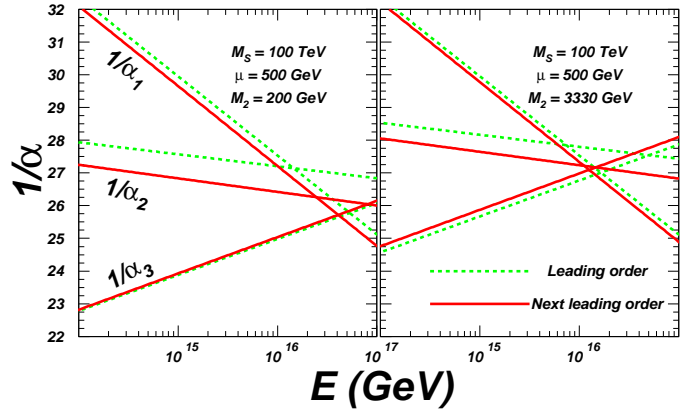}
\end{center}
\vspace{-0.5cm}
\caption{The RGE running of the three gauge couplings (we only show
the region of $E>10^{14}$ GeV).
The dashed lines (green) denote the one-loop results while the solid lines (red)
denote the two-loop results.}
 \label{fig1}
\end{figure}

The RGE running of the three gauge couplings for some benchmark points in the
parameter space is displayed in Fig.\ref{fig1}, where we fix
$M_S=100 {\rm ~TeV},\mu=500 {\rm ~GeV},\tan\beta=10$ and vary $M_2$ from 200 GeV
to 3.33 TeV.
To illustrate if the three gauge couplings can really merge at a high scale,
we only show the running region of $E>10^{14}$ GeV in this figure.
In fact, we found that the two-loop RGEs change $g_2$ coupling more sizably
than $g_1$ and $g_3$.
We can see from this figure that gauge coupling unification prefers a relatively large
gaugino mass.

\begin{figure}[htbp]
\begin{center}
\includegraphics[width=3.5in]{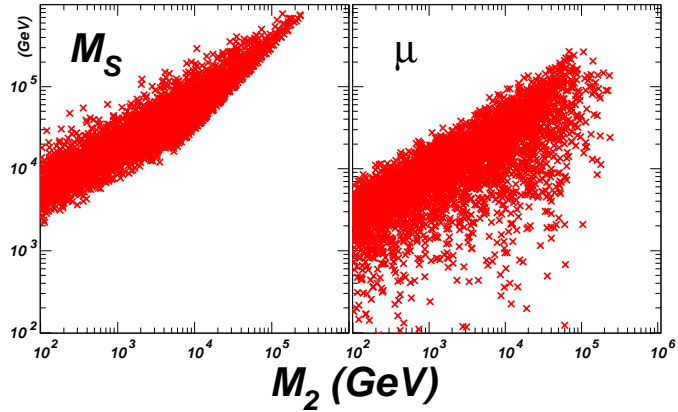}
\end{center}
\vspace{-0.7cm}
\caption{The scatter plots of the parameter space
with the gauge coupling unification requirement.}
\label{fig2}
\end{figure}

With a random scan over the parameter space ($0<M_2,\mu<M_S\leq 10^{13}{\rm GeV}$) for $1<\tan\beta<50$ under the gauge coupling
unification requirement, we obtain the results shown in Fig. \ref{fig2}.
The sharp edge within the figures corresponds to the constraints $M_S>M_3$ in the split SUSY.
From the left panel we can find
an upper bound for $M_S$, which is about $10^6$ GeV
(since split-SUSY requires $M_S\gg M_{\tl{g}_i}$, we can also obtain an upper bound
on $M_2$ correspondingly).
From the right panel we can find upper
limits for $\mu$ and $M_2$, which are around 100 TeV, independent of
the $M_S$ value.

\begin{figure}[htbp]
\begin{center}
\includegraphics[width=3.5in]{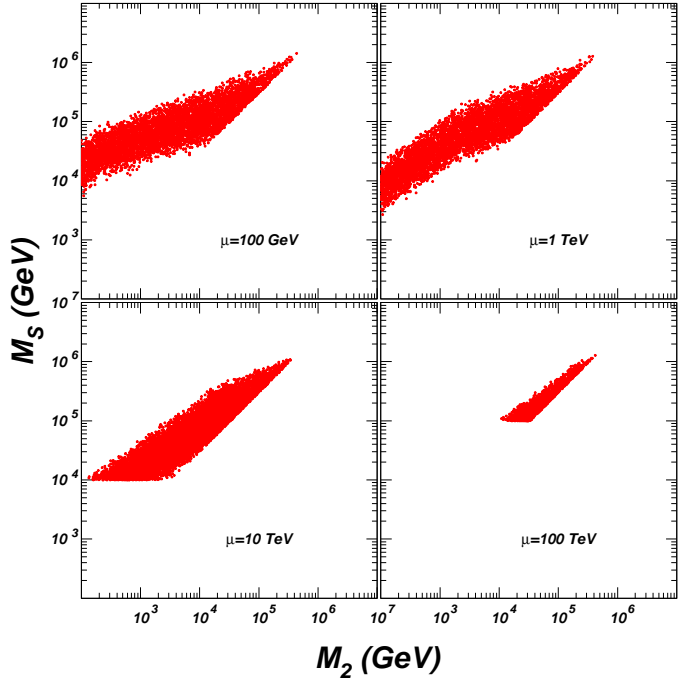}
\end{center}
\vspace{-0.7cm}
\caption{Same as Fig.2, but showing $M_2$ versus $M_S$ for
fixed $\mu$.}
\label{fig3}
\end{figure}

We also scan the parameter space of ($M_2$, $M_S$) with a fixed value of $\mu$
and display the results in Fig.\ref{fig3}.
We can see that the gauge coupling unification imposes a lower bound
on $M_S$, which is 5 TeV for a small $\mu$ value.
It is also interesting to note that a lower bound for $M_2$
exists for a large $\mu$ value.
However, when $\mu$ turns small, the lower bound for $M_2$ is relaxed.

Note that on the plane of  ($M_2$, $M_S$) the  gauge coupling unification
requirement gives a region instead of a line. The reason is that
some uncertainties are involved in gauge coupling unification requirement.
The first uncertainty comes from the measured gauge couplings at $M_Z$ scale
and in our calculation we considered the $3\sigma$ range of $\alpha_s(M_Z)$.
The second uncertainty is that the merging
of three gauge couplings at some GUT scale is not 'exact' numerically
(in our analysis we require the difference between any two gauge couplings
to be smaller than 0.005 while the gauge coupling strength is about 0.68).

 We should give a brief comment on the role of parameter $\tan\beta$ in the gauge coupling unification. Naively, $\tan\beta$ does not appear explicitly in the two-loop
 gauge coupling beta functions. However, $\tan\beta$ can affect the gauge coupling RGE running by showing itself in the
yukawa couplings and the gaugino couplings $\tl{g},\tl{g}^\pr$. Numerical studies indicates that the unification is not sensitive to the choice of $\tan\beta$.
The parameter $M_i,\mu$, which define the thresholds of gauginos and higgsino, can also affect the gauge coupling unification by changing the value of beta functions.

\section{\label{dm}Dark matter in split-SUSY}
In split-SUSY the lightest neutralino $\tilde{\chi}_0$ is proposed to be
the Weakly Interacting Massive Particle (WIMP) dark matter candidate.
We now check the dark matter issue in split-SUSY, using the latest
relic density data from Planck and the direct detection limits from XENON100,LUX as well as the future Xeon1T.

 We use the package DarkSUSY \cite{darksusy} to scan the parameter
space of split-SUSY in the ranges:
\begin{eqnarray}
1<\tan\beta<50, ~~0~<(M_2,~\mu)< M_S.
\end{eqnarray}
  In order to use DarkSUSY to calculate the relic density of dark matter in split susy scenario,
we use the fact that the effects of heavy sfermions and heavy higgs almost entirely decouple when $M_S=M_A>5 {\rm TeV}\cite{FWY}$.
So in our numerical study, we single out the points which satisfy the GUT constraints (as that in previous section) and then set $M_S=M_A=10 {\rm TeV}$ in DarkSUSY to carry
out dark matter related numerical calculations for such survived points.

In our scan we take into account the current dark matter and collider constraints:
\bit
\item[(1)] We use the lightest neutralino $\tilde{\chi}_1^0$ to account for the
Planck measured dark matter relic density $\Omega_{DM} = 0.1199\pm 0.0027$ \cite{planck}
(in combination with the WMAP data \cite{wmap});
\item[(2)] The LEP lower bounds on neutralino and charginos, including the invisible decay of $Z$-boson;
 For LEP experiments, the most stringent constraints come from the chargino mass and the invisible $Z$-boson decay.
 We require that $m_{\tl{\chi}^\pm}> 103 {\rm GeV}$ and the invisible decay width $\Gamma(Z\ra \tl{\chi}_0\tl{\chi}_0)<1.71~{\rm MeV}$,
 which is consistent with the $2\sigma$ precision EW measurement result: $\Gamma^{non-SM}_{inv}< 2.0~{\rm MeV}$.

\item[(3)] The precision electroweak measurements;

Indirect constraints from electroweak precision observables such as $\rho_l, \sin^2\theta_{eff}^l$
 and $M_W$ or their combinations (oblique parameters $S,T,U$)\cite{obliquep}. We require the oblique parameters to be compatible with
the LEP/SLD data at 2$\sigma$ confidence level \cite{stuconstraints}. We compute these observables with the formula presented in \cite{stuformula}.

\item[(4)] The combined mass range for the Higgs boson: $123 {\rm GeV}<M_h <127 {\rm GeV}$
from ATLAS and CMS collaborations of LHC.

  In split-SUSY due to large $M_S$, $\log({m_{\tl{f}}^2/m_t^2})\gg 1$ will spoil the convergence of the traditional loop expansion in
evaluating the SUSY effects of Higgs boson self-energy. So in order to calculate mass of the SM-like Higgs boson,
we use the RGE improved effective potential\cite{effpotential}. This computation method is employed in the NMSSMTools package\cite{nmssmtools}.
This package can be applied to the MSSM cases by setting $\la=\ka\ra 0$  so that the MSSM phenomenology is
recovered.
\eit

We calculate the spin-independent (SI) dark matter-nucleon scattering rate with the relevant
parameters chosen as \cite{Djoudi,Carena,Hisano:2010ct}:
$f_{T_u}^{(p)} =0.023$, $f_{T_d}^{(p)} = 0.032$,
$f_{T_u}^{(n)} = 0.017$,
$f_{T_d}^{(n)} = 0.041$ and $f_{T_s}^{(p)} = f_{T_s}^{(n)} = 0.020$.
In our calculation of the scattering rate, we take into account all the
contributions known so far (including QCD corrections).
For $f_{T_s}$ we take a more reliable value from the recent lattice simulation \cite{lattice}.

\begin{figure}[htbp]
\begin{center}
\includegraphics[width=3.5in]{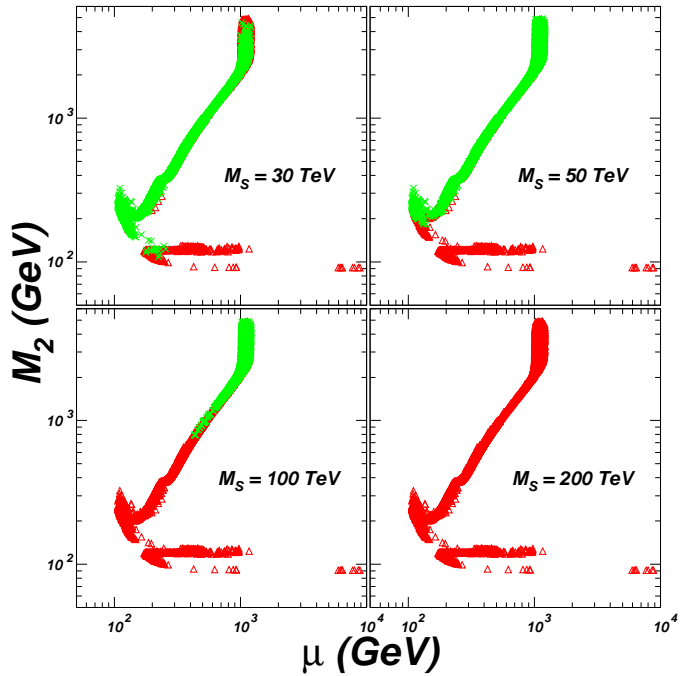}
\end{center}
\vspace{-0.5cm}
\caption{The scatter plots of the parameter space for $\mu>0$ satisfying constraints (1-4)
including dark matter relic density.
The triangles (red) cannot achieve the gauge coupling unification.}
\label{fig4}
\end{figure}

\begin{figure}[htbp]
\begin{center}
\includegraphics[width=3.5in]{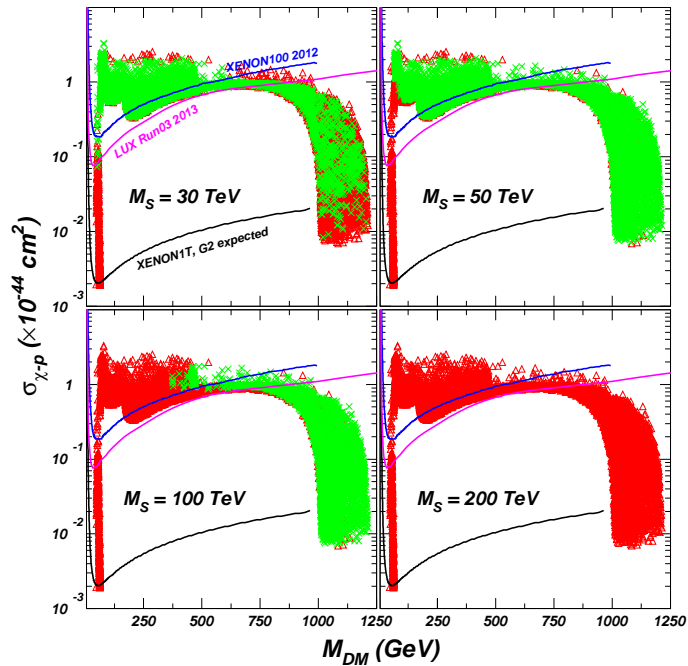}
\end{center}
\vspace{-0.5cm}
\caption{Same as Fig.4, but showing the spin-independent cross section
of dark matter scattering off the nucleon. The curves denote the limits from
 LUX \cite{lux} and XENON100 as well as the future XENON-1T sensitivity. }
\label{fig5}
\end{figure}

In figs.\ref{fig4} and \ref{fig5}, we show the scatter plots of the parameter space
satisfying constraints (1)-(4) with positive $\mu$. In the allowed parameter space, some samples
cannot achieve the gauge coupling unification, which are marked out with red color in these figures.
From fig.4, we can see that all the parameter space satisfying constraints (1-4) are excluded by GUT constraints for $M_S\gtrsim200$ TeV.

We see that the current LUX\cite{lux} and XENON100 direct detection limits are quite stringent for
split-SUSY, which can exclude a large part of
the parameter space allowed by other constraints
including the dark matter relic density.
  Note that a strip corresponding to a dark matter mass range from 1.0 TeV to 1.3 TeV
can survive the combined constraints of GUT and dark matter direct detection for $M_S\lesssim 200$ TeV.
From a careful analysis we found that this strip of parameter space gives
a higgsino-like dark matter.
Outside this strip (i.e. for a dark matter mass below 1 TeV), the survived parameter space can be fully covered by the future XENON-1T experiment.
 In fact, the vast majority of such survived parameter spaces had already been excluded by LUX.

\begin{figure}[htbp]
\begin{center}
\includegraphics[width=3.5in]{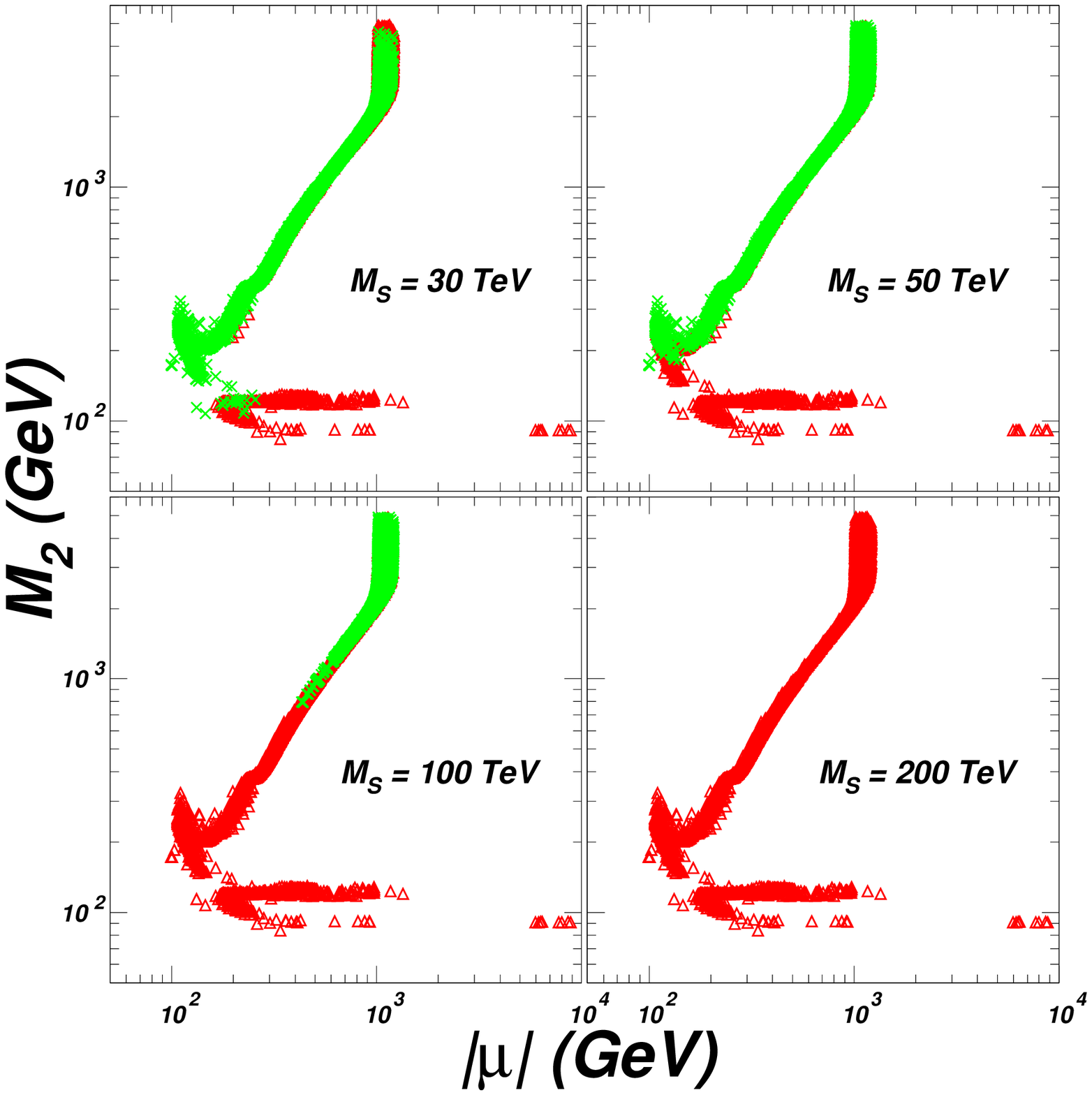}
\end{center}
\vspace{-0.5cm}
\caption{Same as Fig.4 for $\mu<0$.}
\label{fig6}
\end{figure}

\begin{figure}[htbp]
\begin{center}
\includegraphics[width=3.5in]{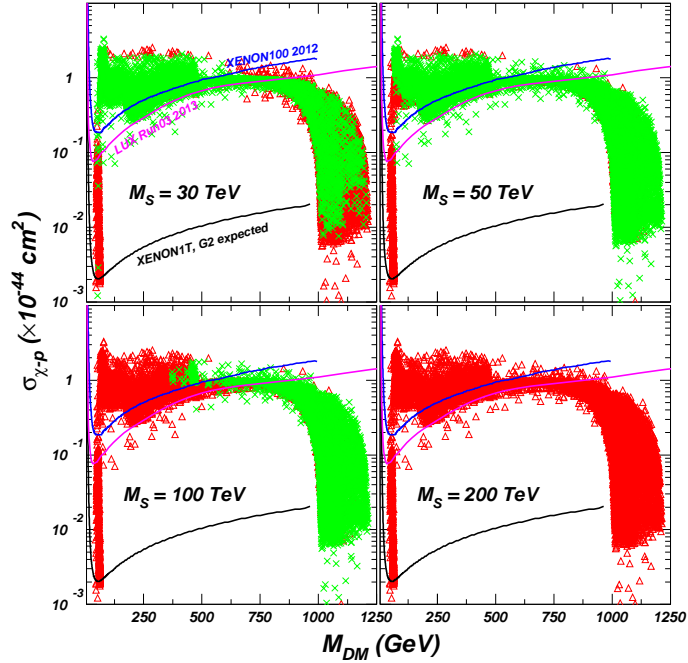}
\end{center}
\vspace{-0.5cm}
\caption{Same as Fig.5 for $\mu<0$.}
\label{fig7}
\end{figure}

 For negative $\mu$, the survived parameter spaces are shown in fig.\ref{fig6} and fig.\ref{fig7}. Our numerical calculations show that in most parameter spaces the results are not very sensitive to the sign of $\mu$.  The minus sign scenario can only revive a very small part of parameter spaces
 which otherwise be excluded in positive $\mu$ scenario. However, unlike the positive $\mu$ scenario, future XENON-1T experiment is necessary to cover all the survived parameter spaces with a dark matter mass below 1 TeV.

So we can conclude that for a dark matter mass below 1 TeV
the split-SUSY under current experimental constraints and
gauge coupling unification requirement can be fully covered by
the future XENON-1T experiment.
\section{\label{conclusion}Conclusion}
We calculated the two-loop beta functions for three gauge couplings
in split-SUSY taking into account all low energy threshold corrections. In split-SUSY scenario
with gaugino mass unification assumption
and a large $M_S$, we find that the gauge coupling unification requires
a lower bound on gaugino mass.
 Combined with the constraints from the dark matter relic density and direct detection limits,
we found that split-SUSY is very restricted and for dark matter mass below 1 TeV
the allowed parameter space can be fully covered by XENON-1T(2017).

We are very grateful to the referee for discussions and comments. This work was supported by the
Natural Science Foundation of China under grant numbers 11105124,11105125,
11275245, 10821504,  11135003, 11005006, 11172008 and Ri-Xin
Foundation of BJUT.
\section*{Appendix A: Boundary Value of the RGE Running}
 We will use the modified minimal subtraction ($\overline{MS}$) scheme in our gauge coupling RGE running.

 Taking into account certain threshold contributions,  the $\overline{MS}$ couplings can be extracted from the standard model input $\al_s(M_Z)=0.1184\pm 0.0007$ by
 \beqa
 \f{\hat{g}_3^2}{4\pi}(M_Z)|_{\overline{MS}}=\f{\al_{s}(M_Z)}{1+\f{\al_s(M_Z)}{2\pi}\f{2}{3}\ln\(\f{m_t}{M_Z}\)}
 \eeqa
  Similarly, we have
\beqa
\hat{\al}_{em}(M_Z)|_{\overline{MS}}=\f{\al_{em}(M_Z)}{1+\f{\al_{em}(M_Z)}{2\pi}\f{16}{9}\ln\(\f{m_t}{M_Z}\)}
\eeqa
with the Standard Model input $\al_{em}^{-1}(M_Z)=127.916\pm 0.015$.

 The exact form of effective weak mixing angle in the modified minimal subtraction $\overline{MS}$ scheme is rather complex and we use the given by PDG\cite{PDG}
 \beqa
 \hat{s}^2\equiv\f{\hat{g}^{\pr2}(M_Z)}{\hat{g}^{\pr2}(M_Z)+\hat{g}^{\pr2}(M_Z)}=0.23116\pm 0.00013.
 \eeqa

 From the top-quark pole mass $M_t|_{pole}=173.5{\rm GeV}$ and taking into account the QCD threshold corrections, one-loop electroweak corrections as well as two-loop ${\cal O}(\al\al_s)$ corrections, the $\overline{MS}$ input for top-yukawa coupling is given by\cite{MStopyukawa}
\beqa
h_t(M_t)&=&0.93587+0.00557\(\f{M_t}{\rm GeV}-173.15\)-0.00003\(\f{M_h}{\rm GeV}-125\)\nn\\
&-&0.00041\(\f{\al_s(M_Z)-0.1184}{0.0007}\).
\eeqa
 In converting the pole top quark mass into $\overline{MS}$ mass, we neglect the subleading possible contributions from gaugino corrections in this stage because of undecided gaugino coupling $\tl{g}_{1d,2d},\tl{g}_{1u,2u}$.

The bottom and tau yukawa couplings at $M_Z$ scale can be similarly extracted from their $\overline{MS}$ or pole mass $m_b(\overline{MS})=4.18{\rm GeV},m_\tau|_{pole}=1.776{\rm GeV}$ followed by RGE running\cite{bernal}
  \beqa
  h_{b}(M_Z)&=&0.024\(1-\f{g_3^2}{8\pi^2}\f{23}{3}\ln\(\f{M_Z}{m_b}\)\)^{12/23}\(1+\f{e^2}{8\pi^2}\f{80}{9}\ln\(\f{M_Z}{m_b}\)\)^{-3/80},\nn\\
  h_{\tau}(M_Z)&=&0.0102\(1-\f{e^2}{4\pi^2}\)\(1+\f{e^2}{8\pi^2}\f{80}{9}\ln\(\f{M_Z}{m_b}\)\)^{-27/80},
  \eeqa

Because of the fact that supersymmetry is not preserved in the $\overline{MS}$ scheme, the boundary conditions appeared in (\ref{condh})
 is valid only in $\overline{DR}$ scheme and will be spoiled in $\overline{MS}$ scheme.
 We know that in case of simple group, the $\overline{MS}$ gauge couplings are related to the $\overline{DR}$ gauge couplings by the relation\cite{martin}
\beqa
g_{\overline{MS}}=g_{\overline{DR}}\[1-\f{g^2}{96\pi^2}C(G)\].
\eeqa
 The relation (\ref{condh}) in $\overline{MS}$ scheme will be changed into
\beqa
\tl{g}_u(M_S)&=&g({M_S})\sin\beta\[1+\f{1}{16\pi^2}\(\f{23}{24}g^2-\f{1}{8}g^{\pr2}\)\],\nn\\
\tl{g}_u^\pr(M_S)&=&g^\pr({M_S})\sin\beta\[1+\f{1}{16\pi^2}\(\f{3}{8}g^2+\f{1}{8}g^{\pr2}\)\],\nn\\
\tl{g}_d(M_S)&=&g({M_S})\sin\beta\[1+\f{1}{16\pi^2}\(\f{23}{24}g^2-\f{1}{8}g^{\pr2}\)\],\nn\\
\tl{g}_d^\pr(M_S)&=&g^\pr({M_S})\sin\beta\[1+\f{1}{16\pi^2}\(\f{3}{8}g^2+\f{1}{8}g^{\pr2}\)\],
\eeqa
at the $M_S$ scale at tree-level. This result agrees with the results in \cite{lodone}( and also agrees with ref.\cite{bernal} if we use the tree-level expression $c^2=g^2/(g^{\pr 2}+g^2)$ to eliminate $g^\pr$).

At one-loop level, the expression changed into \cite{lodone}
\beqa
\f{\tl{g}_u(M_S)}{g({M_S})\sin\beta}&=&1+\f{1}{16\pi^2}\[\f{23}{24}g^2-\f{1}{8}g^{\pr2}+
\f{7}{16}\cos^2 \beta~g^{\pr 2} -\(\f{11}{16}\cos^2 \beta+\f{13}{8}\)g^{2}+\(\f{3}{4\sin^2\beta} +\f{3}{2}\)h_t^2\]£¬\nn\\
\f{\tl{g}_u^\pr(M_S)}{g^\pr({M_S})\sin\beta}&=& 1+\f{1}{16\pi^2}\[\f{3}{8}g^2+\f{1}{8}g^{\pr2}+\f{21}{16}\cos^2 \beta~g^{2} +\(\f{7}{16}\cos^2 \beta-\f{21}{8}\)g^{\pr 2}+\(\f{3}{4\sin^2\beta} +\f{3}{2}\)h_t^2\],\nn\\
\f{\tl{g}_d(M_S)}{g({M_S})\sin\beta}&=& 1+\f{1}{16\pi^2}\[\f{23}{24}g^2-\f{1}{8}g^{\pr2}+
\f{7}{16}\sin^2 \beta~g^{\pr 2} -\(\f{11}{16}\sin^2 \beta+\f{13}{8}\)g^{2}+\f{3}{2}h_t^2\],\nn\\
\f{\tl{g}_d^\pr(M_S)}{g^\pr({M_S})\sin\beta}&=&1+\f{1}{16\pi^2}\[\f{3}{8}g^2+\f{1}{8}g^{\pr2}+\f{21}{16}\sin^2 \beta~g^{2} +\(\f{7}{16}\sin^2 \beta-\f{21}{8}\)g^{\pr 2}+\f{3}{2}h_t^2\].
\eeqa
with proper normalization $g^{\pr}=\sqrt{3/5}g_1$. Because such boundary conditions are given at the $M_S$ scale while other inputs are given at the weak scale $M_Z$, iterative procedure is necessary in the numerical studies.
\section*{Appendix B: Two-Loop RGE for Gauge Couplings in Split Supersymmetry}

The 2-loop RGE for $SU(3)_c, SU(2)_L, U(1)_Y$ gauge couplings ($g_3,g_2,g_1$, respectively) are given by
\beqa
\f{d}{d\ln E} g_i=\f{b_i}{(4\pi)^2}g_i^3+\f{g_i^3}{(4\pi)^4}\[\sum\limits_{j} B_{ij}g_j^2-\sum\limits_{a=u,d,e}d_i^a Tr(h^{a\da} h^a)-d_W(\tl{g}_u^2+\tl{g}_d^2)-d_B(\tl{g}_u^{\pr 2}+\tl{g}_d^{\pr 2})\],\nn
\eeqa
with the $U(1)_Y$ normalization $g_1^2=\f{5}{3}(g_Y)^2$ and the relevant coefficients in Table \ref{tab1},\ref{tab2},\ref{tab3},\ref{tab4}.

 The one-loop RGE for Yukawa couplings below the $M_S$ scale can be written as
\beqa
16\pi^2 \f{d}{dt} h^u&=& h^u\[-3c_i^u g_i^2+c_{T}^u T+ c_{S_1}^u S_1+ c_{S_2}^u S_2+\f{3}{2}\(h^{u\da}h^u-h^{d\da}h^d\)\],\nn\\
16\pi^2 \f{d}{dt} h^d&=& h^d\[-3c_i^d g_i^2+c_{T}^d T+c_{S_1}^d S_1+ c_{S_2}^d S_2+\f{3}{2}\(h^{d\da}h^d-h^{u\da}h^u\)\],\nn\\
16\pi^2 \f{d}{dt} h^e&=& h^e\[-3c_i^e g_i^2+c_{T}^e TT+c_{S_1}^e S_1+c_{S_2}^e S_2+\f{3}{2}h^{e\da}h^e\],\nn\\
\eeqa
with
\beqa
T&=&Tr(3h^{u\da}h^u+3h^{d\da}h^d+h^{e\da}h^e),~S_1=\f{1}{2}\[(\tl{g}_u^\pr)^2+(\tl{g}_d^\pr)^2\],~S_2=\f{3}{2}\(\tl{g}_u^2+\tl{g}_d^2\),\nn
\eeqa
The relevant coefficients in different scenarios can be found in Table \ref{tab5},\ref{tab6},\ref{tab7}.

Upon $M_S$, we recover the MSSM result and the one-loop RGE for yukawa-type interactions in the superpotential are
\beqa
16\pi^2 \f{d}{dt} \la^u&=& \la^u\[-2c_i^u g_i^2+3Tr(\la^{u\da}\la^u)+3\la^{u\da}\la^u+\la^{d\da}\la^d\],\nn\\
16\pi^2 \f{d}{dt} \la^d&=& \la^d\[-2c_i^d g_i^2+Tr(3\la^{d\da}\la^d+\la^{e\da}\la^e)+\la^{u\da}\la^u+3\la^{d\da}\la^d\],\nn\\
16\pi^2 \f{d}{dt} \la^e&=& \la^e\[-2c_i^e g_i^2+Tr(3\la^{d\da}\la^d+\la^{e\da}\la^e)+3\la^{e\da}\la^e\],\nn\\
\eeqa
with
\beqa
c_i^u=(\f{13}{30},\f{3}{2},\f{8}{3}),~c_i^d=(\f{7}{30},\f{3}{2},\f{8}{3}),~c_i^e=(\f{9}{10},\f{3}{2},0).\nn
\eeqa

The gaugino coupling RGE (upon gaugino, higgsino thresholds and below $M_S$) can be written as
\beqa
16\pi^2\f{d }{dt} \tl{g}_{u}&=&-3\tl{g}_u c_i^u g_i^2+\f{5}{4}\tl{g}_u^3-\f{1}{2}\tl{g}_u\tl{g}_d^2+\f{1}{4}\tl{g}_u\tl{g}_u^{\pr 2}+\tl{g}_d\tl{g}_d^\pr\tl{g}_u^\pr+\tl{g}_u (T+c_{S_1} S_1+c_{S_2}S_2),\nn\\
16\pi^2\f{d }{dt} \tl{g}_{d}&=&-3\tl{g}_d c_i^d g_i^2+\f{5}{4}\tl{g}_d^3-\f{1}{2}\tl{g}_d\tl{g}_u^2+\f{1}{4}\tl{g}_d\tl{g}_d^{\pr 2}+\tl{g}_u\tl{g}_u^\pr\tl{g}_d^\pr+\tl{g}_d (T+c_{S_1} S_1+c_{S_2}S_2),\nn\\
16\pi^2\f{d }{dt} \tl{g}_{u}^\pr&=&-3\tl{g}_u^\pr \tl{c}_i^u g_i^2+\f{3}{4}\tl{g}_u^{\pr 3}+\f{3}{2}\tl{g}^\pr_u\tl{g}_d^{\pr 2}+\f{3}{4}\tl{g}_u^\pr\tl{g}_u^{2}+3\tl{g}_d^\pr\tl{g}_d\tl{g}_u+\tl{g}_u^\pr (T+c_{S_1} S_1+c_{S_2}S_2),\nn\\
16\pi^2\f{d }{dt} \tl{g}_{d}^\pr&=&-3\tl{g}_d^\pr \tl{c}_i^d g_i^2+\f{3}{4}\tl{g}_d^{\pr 3}+\f{3}{2}\tl{g}^{\pr}_ d\tl{g}^{\pr 2}_u+\f{3}{4}\tl{g}^{\pr}_d
\tl{g}_d^{2}+3\tl{g}_u^\pr\tl{g}_u \tl{g}_d+\tl{g}_d^\pr (T+c_{S_1} S_1+c_{S_2}S_2),~
\eeqa
with the coefficient
\beqa
c_i^{u,d}=(\f{3}{20},\f{11}{4},0),~~\tl{c}_i^{u,d}=(\f{3}{20},\f{3}{4},0),~~c_{S_1}=c_{S_2}=1,
\eeqa
and the boundary value at $M_S$ scale
 \beqa
 \tl{g}_u(M_S)&=&g_2(M_S)\sin\beta,~~\tl{g}_d(M_S)=g_2(M_S)\cos\beta, \nn\\
 \tl{g}_u^\pr(M_S)&=&g_1(M_S)\sin\beta,~~\tl{g}_d^\pr(M_S)=g_1(M_S)\cos\beta.
\eeqa
Below $M_2$, we can decoupling the effect of wino by setting $\tl{g}_u=\tl{g}_d=0$. Blow $M_1$, the effect of bino can be decoupled by setting $ \tl{g}_{u}^\pr= \tl{g}_{d}^\pr=0$. Below $\mu$, these gaugino interactions will decouple.

\small
\begin{table}
\caption{The coefficients in two-loop gauge coupling RGE with $M_3<\mu< M_S$.}
\label{tab1}
\begin{tabular}{|c|c|c|c|c|}
\hline
$E$& $b_i$& $ B_{ij}$ & $(d^u_i, d^d_i, d^e_i)$& $(d^W_i,d^B_i)$ \\
\hline $[M_Z,M_2]$ & $\(\bea{c}41/10\\-19/6\\-7\eea\)$&$\(\bea{ccc}\f{199}{50}&\f{27}{10}&\f{44}{5}\\\f{9}{10}&\f{35}{6}&12\\\f{11}{10}&\f{9}{2}&-26\eea\)$  &    $\(\bea{ccc}\f{17}{10}&\f{1}{2}&\f{3}{2}\\\f{3}{2}&\f{3}{2}&\f{1}{2}\\ 2
 & 2&0\eea \)$ & $  \(\bea{cc}0&0\\0&0\\0&0\eea\) $ \\
\hline $[M_2,M_3]$ &$ \(\bea{c}~41/10\\-11/6\\-7\eea\) $
 &$\(\bea{ccc}\f{199}{50}&\f{27}{10}&\f{44}{5}\\\f{9}{10}&\f{163}{6}&12\\\f{11}{10}&\f{9}{2}&-26\eea\)$
 &$\(\bea{ccc}\f{17}{10}&\f{1}{2}&\f{3}{2}\\\f{3}{2}&\f{3}{2}&\f{1}{2}\\ 2
 & 2&0\eea \)$&$\( \bea{cc}\f{9}{20}&\f{3}{20}\\0&0\\0&0\eea\) $ \\
\hline $[M_3,\mu]$&$ \(\bea{c}~41/10\\-11/6\\-5\eea\) $
 &$\(\bea{ccc}\f{199}{50}&\f{27}{10}&\f{44}{5}\\\f{9}{10}&\f{163}{6}&12\\\f{11}{10}&\f{9}{2}&22\eea\) $&
 $\(\bea{ccc}\f{17}{10}&\f{1}{2}&\f{3}{2}\\\f{3}{2}&\f{3}{2}&\f{1}{2}\\ 2
 & 2&0\eea \)$
 &$ \(\bea{cc}\f{9}{20}&\f{3}{20}\\\f{11}{4}&\f{1}{4}\\0&0\eea \)$ \\
\hline $[\mu, M_S]$&$ \(\bea{c}9/2\\-7/6\\-5\eea\) $
 &$ \(\bea{ccc}\f{104}{25}&\f{18}{5}&\f{44}{5}\\\f{6}{5}&\f{106}{3}&12\\\f{11}{10}&\f{9}{2}&22\eea\) $&
 $\(\bea{ccc}\f{17}{10}&\f{1}{2}&\f{3}{2}\\\f{3}{2}&\f{3}{2}&\f{1}{2}\\ 2
 & 2&0\eea\) $& $ \(\bea{cc}\f{9}{20}
 &\f{3}{20}\\\f{11}{4}&\f{1}{4}\\0&0\eea\) $ \\
\hline $[M_S,M_{U}]$&$ \(\bea{c}\f{33}{5}\\1\\-3\eea\) $&$ \(\bea{ccc}\f{199}{25}&\f{27}{5}&\f{88}{5}\\\f{9}{5}&25&24\\\f{11}{5}&9&14\eea\)$&
$\(\bea{ccc}\f{26}{5}&\f{14}{5}&\f{18}{5}\\6 &6&2\\ 4
 & 4&0\eea\) $&$ \(\bea{cc}0&0\\0&0\\0&0\eea\) $ \\
\hline
\end{tabular}
\end{table}
\normalsize

\small
\begin{table}[htbp]
\caption{The coefficients in two-loop gauge coupling RGE with $M_2<\mu< M_3$.}
\label{tab2}
\begin{tabular}{|c|c|c|c|c|}
\hline
$E$& $b_i$& $ B_{ij}$ & $(d^u_i, d^d_i, d^e_i)$& $(d^W_i,d^B_i)$ \\
\hline $[M_Z,M_2]$ & $\(\bea{c}41/10\\-19/6\\-7\eea\)$&$\(\bea{ccc}\f{199}{50}&\f{27}{10}&\f{44}{5}\\\f{9}{10}&\f{35}{6}&12\\\f{11}{10}&\f{9}{2}&-26\eea\)$  &    $\(\bea{ccc}\f{17}{10}&\f{1}{2}&\f{3}{2}\\\f{3}{2}&\f{3}{2}&\f{1}{2}\\ 2
 & 2&0\eea \)$ & $  \(\bea{cc}0&0\\0&0\\0&0\eea\) $ \\
\hline $[M_2,\mu]$ &$ \(\bea{c}~41/10\\-11/6\\-7\eea\) $
 &$\(\bea{ccc}\f{199}{50}&\f{27}{10}&\f{44}{5}\\\f{9}{10}&\f{163}{6}&12\\\f{11}{10}&\f{9}{2}&-26\eea\)$
 &$\(\bea{ccc}\f{17}{10}&\f{1}{2}&\f{3}{2}\\\f{3}{2}&\f{3}{2}&\f{1}{2}\\ 2
 & 2&0\eea \)$&$\( \bea{cc}\f{9}{20}&\f{3}{20}\\0&0\\0&0\eea\) $ \\
\hline $[\mu,M_3]$&$ \(\bea{c}9/2\\-7/6\\-7\eea\) $
 &$\(\bea{ccc}\f{199}{50}&\f{27}{10}&\f{44}{5}\\\f{9}{10}&\f{163}{6}&12\\\f{11}{10}&\f{9}{2}&22\eea\) $&
 $\(\bea{ccc}\f{17}{10}&\f{1}{2}&\f{3}{2}\\\f{3}{2}&\f{3}{2}&\f{1}{2}\\ 2
 & 2&0\eea \)$
 &$ \(\bea{cc}\f{9}{20}&\f{3}{20}\\\f{11}{4}&\f{1}{4}\\0&0\eea \)$ \\
\hline $[M_3, M_S]$&$ \(\bea{c}9/2\\-7/6\\-5\eea\) $
 &$ \(\bea{ccc}\f{104}{25}&\f{18}{5}&\f{44}{5}\\\f{6}{5}&\f{106}{3}&12\\\f{11}{10}&\f{9}{2}&22\eea\) $&
 $\(\bea{ccc}\f{17}{10}&\f{1}{2}&\f{3}{2}\\\f{3}{2}&\f{3}{2}&\f{1}{2}\\ 2
 & 2&0\eea\) $& $ \(\bea{cc}\f{9}{20}
 &\f{3}{20}\\\f{11}{4}&\f{1}{4}\\0&0\eea\) $ \\
\hline $[M_S,M_{U}]$&$ \(\bea{c}\f{33}{5}\\1\\-3\eea\) $&$ \(\bea{ccc}\f{199}{25}&\f{27}{5}&\f{88}{5}\\\f{9}{5}&25&24\\\f{11}{5}&9&14\eea\)$&
$\(\bea{ccc}\f{26}{5}&\f{14}{5}&\f{18}{5}\\6 &6&2\\ 4
 & 4&0\eea\) $&$ \(\bea{cc}0&0\\0&0\\0&0\eea\) $ \\
\hline
\end{tabular}
\end{table}
\normalsize

\small
\begin{table}[htbp]
\caption{The coefficients in two-loop gauge coupling RGE with $M_1<\mu< M_2$.}
\label{tab3}
\begin{tabular}{|c|c|c|c|c|}
\hline
$E$& $b_i$& $ B_{ij}$ & $(d^u_i, d^d_i, d^e_i)$& $(d^W_i,d^B_i)$ \\
\hline $[M_Z,\mu]$ & $\(\bea{c}41/10\\-19/6\\-7\eea\)$&$\(\bea{ccc}\f{199}{50}&\f{27}{10}&\f{44}{5}\\\f{9}{10}&\f{35}{6}&12\\\f{11}{10}&\f{9}{2}&-26\eea\)$  &    $\(\bea{ccc}\f{17}{10}&\f{1}{2}&\f{3}{2}\\\f{3}{2}&\f{3}{2}&\f{1}{2}\\ 2
 & 2&0\eea \)$ & $  \(\bea{cc}0&0\\0&0\\0&0\eea\) $ \\
\hline $[\mu,M_2]$ &$ \(\bea{c}~9/2\\-15/6\\-7\eea\) $
 &$\(\bea{ccc}\f{199}{50}&\f{27}{10}&\f{44}{5}\\\f{9}{10}&\f{163}{6}&12\\\f{11}{10}&\f{9}{2}&-26\eea\)$
 &$\(\bea{ccc}\f{17}{10}&\f{1}{2}&\f{3}{2}\\\f{3}{2}&\f{3}{2}&\f{1}{2}\\ 2
 & 2&0\eea \)$&$\( \bea{cc}\f{9}{20}&\f{3}{20}\\0&0\\0&0\eea\) $ \\
\hline $[M_2,M_3]$&$ \(\bea{c}9/2\\-7/6\\-7\eea\) $
 &$\(\bea{ccc}\f{104}{25}&\f{18}{5}&\f{44}{5}\\\f{6}{5}&\f{106}{3}&12\\\f{11}{10}&\f{9}{2}&-26\eea\)$&
 $\(\bea{ccc}\f{17}{10}&\f{1}{2}&\f{3}{2}\\\f{3}{2}&\f{3}{2}&\f{1}{2}\\ 2
 & 2&0\eea \)$
 &$ \(\bea{cc}\f{9}{20}&\f{3}{20}\\\f{11}{4}&\f{1}{4}\\0&0\eea \)$ \\
\hline $[M_3, M_S]$&$ \(\bea{c}9/2\\-7/6\\-5\eea\) $
 &$ \(\bea{ccc}\f{104}{25}&\f{18}{5}&\f{44}{5}\\\f{6}{5}&\f{106}{3}&12\\\f{11}{10}&\f{9}{2}&22\eea\) $&
 $\(\bea{ccc}\f{17}{10}&\f{1}{2}&\f{3}{2}\\\f{3}{2}&\f{3}{2}&\f{1}{2}\\ 2
 & 2&0\eea\) $& $ \(\bea{cc}\f{9}{20}
 &\f{3}{20}\\\f{11}{4}&\f{1}{4}\\0&0\eea\) $ \\
\hline $[M_S,M_{U}]$&$ \(\bea{c}\f{33}{5}\\1\\-3\eea\) $&$ \(\bea{ccc}\f{199}{25}&\f{27}{5}&\f{88}{5}\\\f{9}{5}&25&24\\\f{11}{5}&9&14\eea\)$&
$\(\bea{ccc}\f{26}{5}&\f{14}{5}&\f{18}{5}\\6 &6&2\\ 4
 & 4&0\eea\) $&$ \(\bea{cc}0&0\\0&0\\0&0\eea\) $ \\
\hline
\end{tabular}
\end{table}
\normalsize


\small
\begin{table}[htbp]
\label{tab4}
\caption{The coefficients in two-loop gauge coupling RGE
with $M_Z<\mu<M_1$.}
\begin{tabular}{|c|c|c|c|c|}
\hline
$E$& $b_i$& $ B_{ij}$ & $(d^u_i, d^d_i, d^e_i)$& $(d^W_i,d^B_i)$ \\
\hline $[M_Z,\mu]$ &$ \(\bea{c}~\f{41}{10}\\-\f{19}{6}\\-7\eea\) $
 & $\(\bea{ccc}\f{199}{50}&\f{27}{10}&\f{44}{5}\\\f{9}{10}
 &\f{35}{6}&12\\\f{11}{10}&\f{9}{2}&-26\eea\) $&
 $ \(\bea{ccc}\f{17}{10}&\f{1}{2}&\f{3}{2}\\\f{3}{2}&\f{3}{2}&\f{1}{2}\\ 2
 & 2&0\eea\) $&$ \(\bea{cc}0&0\\0&0\\0&0\eea\) $ \\
\hline $[\mu,M_1]$ &$ \(\bea{c}~\f{9}{2}\\-\f{15}{6}\\-7\eea\) $
 &$\( \bea{ccc}\f{104}{25}&\f{18}{5}&\f{44}{5}\\\f{6}{5}
 &\f{42}{3}&12\\\f{11}{10}&\f{9}{2}&-26\eea\) $
 &$ \(\bea{ccc}\f{17}{10}&\f{1}{2}&\f{3}{2}\\\f{3}{2}&\f{3}{2}&\f{1}{2}\\ 2
 & 2&0\eea\) $&$ \(\bea{cc}0&0\\0&0\\0&0\eea \)$ \\
\hline $[M_1,M_2]$ &$\( \bea{c}~\f{9}{2}\\-\f{15}{6}\\-7\eea\) $
 &$ \(\bea{ccc}\f{104}{25}&\f{18}{5}&\f{44}{5}\\\f{6}{5}
 &\f{42}{3}&12\\\f{11}{10}&\f{9}{2}&-26\eea\) $
 &$ \(\bea{ccc}\f{17}{10}&\f{1}{2}&\f{3}{2}\\\f{3}{2}
 &\f{3}{2}&\f{1}{2}\\ 2& 2&0\eea \)$&$ \(\bea{cc}\f{9}{20}&\f{3}{20}\\0&0\\0&0\eea \)$ \\
\hline $[M_2,M_3]$&$ \(\bea{c}~\f{9}{2}\\-\f{7}{6}\\-7\eea\) $
 &$ \(\bea{ccc}\f{104}{25}&\f{18}{5}&\f{44}{5}\\\f{6}{5}&\f{106}{3}
 &12\\\f{11}{10}&\f{9}{2}&-26\eea \)$&$ \(\bea{ccc}\f{17}{10}&\f{1}{2}
 &\f{3}{2}\\\f{3}{2}&\f{3}{2}&\f{1}{2}\\ 2& 2&0\eea\) $
 &$\( \bea{cc}\f{9}{20}&\f{3}{20}\\\f{11}{4}&\f{1}{4}\\0&0\eea\) $ \\
\hline $[M_3, M_S]$&$ \(\bea{c}~\f{9}{2}\\-\f{7}{6}\\-5\eea\) $
 &$ \(\bea{ccc}\f{104}{25}&\f{18}{5}&\f{44}{5}\\\f{6}{5}
 &\f{106}{3}&12\\\f{11}{10}&\f{9}{2}&22\eea\) $
 &$ \(\bea{ccc}\f{17}{10}&\f{1}{2}&\f{3}{2}\\\f{3}{2}
 &\f{3}{2}&\f{1}{2}\\ 2& 2&0\eea\) $&$\( \bea{cc}\f{9}{20}
 &\f{3}{20}\\\f{11}{4}&\f{1}{4}\\0&0\eea\) $ \\
\hline $[M_S,M_{U}]$&$\( \bea{c}~\f{33}{5}\\~1\\-3\eea\) $
 &$\( \bea{ccc}\f{199}{25}&\f{27}{5}&\f{88}{5}\\\f{9}{5}
 &25&24\\\f{11}{5}&9&14\eea\) $&
 $ \(\bea{ccc}\f{26}{5}&\f{14}{5}&\f{18}{5}\\6&6&2\\ 4
 & 4&0\eea\) $&$ \(\bea{cc}0&0\\0&0\\0&0\eea \)$ \\
\hline
\end{tabular}
\end{table}
\normalsize

\small
\begin{table}[htbp]
\caption{The coefficients in the one-loop yukawa couplings
in case $M_3<\mu<M_S$ and $M_2<\mu<M_3$}.
\label{tab5}
\begin{tabular}{|c|c|c|}
\hline
$E$& $[M_Z,\mu]$&$[\mu, M_S]$ \\
\hline $\(\bea{c} c^u_i\\ c^d_i\\ c^e_i \eea\)$ & $\(\bea{ccc}~\f{17}{60}&\f{3}{4}&\f{8}{3}\\\f{1}{12}&\f{3}{4}&\f{8}{3}\\\f{3}{4}&\f{3}{4}& 0\eea\)$
&$\(\bea{ccc}~\f{17}{60}&\f{3}{4}&\f{8}{3}\\\f{1}{12}&\f{3}{4}&\f{8}{3}\\\f{3}{4}&\f{3}{4}& 0\eea\)$ \\
\hline
$\(\bea{ccc} c^u_{T}&c^u_{S_1}&c^u_{S_2}\\ c^d_{T}&c^d_{S_1}&c^d_{S_1}\\ c^e_{T}&c^e_{S_1}&c^e_{S_2}\eea\)$&$\(\bea{ccc}1&0&0\\1&0&0\\1&0&0\eea\)$&$\(\bea{ccc}1&1&1\\1&1&1\\1&1&1\eea\)$\\
\hline
\end{tabular}
\end{table}
\normalsize

\small
\begin{table}
\caption{The coefficients in the one-loop yukawa couplings
in case $M_1<\mu<M_2$}.
\label{tab6}
\begin{tabular}{|c|c|c|c|}
\hline
$E$& $[M_Z,\mu]$&$[\mu, M_2]$&$[M_2,M_S]$ \\
\hline $\(\bea{c} c^u_i\\ c^d_i\\ c^e_i \eea\)$ & $\(\bea{ccc}~\f{17}{60}&\f{3}{4}&\f{8}{3}\\\f{1}{12}&\f{3}{4}&\f{8}{3}\\\f{3}{4}&\f{3}{4}& 0\eea\)$
&$\(\bea{ccc}~\f{17}{60}&\f{3}{4}&\f{8}{3}\\\f{1}{12}&\f{3}{4}&\f{8}{3}\\\f{3}{4}&\f{3}{4}& 0\eea\)$&$\(\bea{ccc}~\f{17}{60}&\f{3}{4}&\f{8}{3}\\\f{1}{12}&\f{3}{4}&\f{8}{3}\\\f{3}{4}&\f{3}{4}& 0\eea\)$ \\
\hline
$\(\bea{ccc} c^u_{T}&c^u_{S_1}&c^u_{S_2}\\ c^d_{T}&c^d_{S_1}&c^d_{S_1}\\ c^e_{T}&c^e_{S_1}&c^e_{S_2}\eea\)$&$\(\bea{ccc}1&0&0\\1&0&0\\1&0&0\eea\)$&$\(\bea{ccc}1&1&0\\1&1&0\\1&1&0\eea\)$
&$\(\bea{ccc}1&1&1\\1&1&1\\1&1&1\eea\)$\\
\hline
\end{tabular}
\end{table}
\normalsize

\small
\begin{table}
\caption{The coefficients in the one-loop yukawa couplings
in case $M_Z<\mu<M_1$}.
\label{tab7}
\begin{tabular}{|c|c|c|c|}
\hline
$E$& $[M_Z,M_1]$&$[M_1, M_2]$&$[M_2,M_S]$ \\
\hline $\(\bea{c} c^u_i\\ c^d_i\\ c^e_i \eea\)$ & $\(\bea{ccc}~\f{17}{60}&\f{3}{4}&\f{8}{3}\\\f{1}{12}&\f{3}{4}&\f{8}{3}\\\f{3}{4}&\f{3}{4}& 0\eea\)$
&$\(\bea{ccc}~\f{17}{60}&\f{3}{4}&\f{8}{3}\\\f{1}{12}&\f{3}{4}&\f{8}{3}\\\f{3}{4}&\f{3}{4}& 0\eea\)$&$\(\bea{ccc}~\f{17}{60}&\f{3}{4}&\f{8}{3}\\\f{1}{12}&\f{3}{4}&\f{8}{3}\\\f{3}{4}&\f{3}{4}& 0\eea\)$ \\
\hline
$\(\bea{ccc} c^u_{T}&c^u_{S_1}&c^u_{S_2}\\ c^d_{T}&c^d_{S_1}&c^d_{S_1}\\ c^e_{T}&c^e_{S_1}&c^e_{S_2}\eea\)$&$\(\bea{ccc}1&0&0\\1&0&0\\1&0&0\eea\)$&$\(\bea{ccc}1&1&0\\1&1&0\\1&1&0\eea\)$
&$\(\bea{ccc}1&1&1\\1&1&1\\1&1&1\eea\)$\\
\hline
\end{tabular}
\end{table}


\normalsize



\begin{thebibliography}{99}
\vspace{-1mm}
\bibitem{atlas} G. Aad et al.(ATLAS Collaboration), Phys. Lett. B710, 49 (2012).

\bibitem{cms} S. Chatrachyan et al.(CMS Collaboration), Phys. Lett.B710, 26 (2012).

\bibitem{diphoton-susy} See, e.g.,
  M.~Carena {\it et al.} JHEP 1203, 014 (2012); JHEP {\bf 1207}, 175 (2012);
  J.~Cao {\it et al.},
  JHEP {\bf 1210}, 079 (2012); JHEP {\bf 1203}, 086 (2012);
  Phys. Lett. B {\bf 710}, 665 (2012);
  U. Ellwanger, JHEP 1203, 044 (2012);
  G. Belanger {\it et al.}, arXiv:1210.1976; arXiv:1208.4952;
  J. F. Gunion, Y. Jiang, S. Kraml, \PRD86, 071702 (2012); \PRL110, 051801 (2013).

 \bibitem{gutmssm}
J.~R.~Ellis, S.~Kelley and D.~V.~Nanopoulos, Phys.\ Lett.\ B {\bf 249}, 441 (1990);
Phys.\ Lett.\ B {\bf 260}, 131 (1991);
U.~Amaldi, W.~de Boer and H.~Furstenau, Phys.\ Lett.\ B {\bf 260}, 447 (1991);
P.~Langacker and M.~X.~Luo, Phys.\ Rev.\ D {\bf 44}, 817 (1991).

\bibitem{su5}
H.~Georgi and S.~L.~Glashow, Phys.\ Rev.\ Lett.\ {\bf 32}, 438 (1974).

\bibitem{cmssm1} G. Aad et al. (ATLAS collaboration), Phys. Lett. B710 (2012) 67 (2011).

                   G. Aad et al. [ATLAS Collaboration], Phys. Rev. D 87 (2013) 012008.

\bibitem{cmssm2}S. Chatrchyan et al. (CMS collaboration), Phys. Rev.Lett. 107 (2011) 221804.

               S. Chatrchyan et al. [CMS Collaboration], J. High Energy Phys. 1210 (2012) 018.

\bibitem{nima} N. Arkani-Hamed, S. Dimopoulos, JHEP 0506 (2005) 073.



\bibitem{giudice1} G.F. Giudice, A. Romaninom, Nucl. Phys. B699, 65 (2004).
\bibitem{giudicenima} N. Arkani-Hamed, S. Dimopoulos, G.F. Giudice, A. Romanino, Nucl. Phys. B709(2005) 3-46.

\bibitem{planck} http://www.sciops.esa.int/SA/PLANCK/docs/Planck 2013 results 16.pdf.

\bibitem{wmap} J. Dunkley et al.[WMAP Collaboration], Astrophys. J. Suppl. 180, 306 (2009).

\bibitem{xenon} E. Aprile et al. [XENON100 Collaboration], Phys. Rev. Lett. 109, 181301 (2012).

\bibitem{cwy}  J. Cao, W. Wang, J. M. Yang, Phys. Lett. B{\bf 706}, 72 (2011).
\bibitem{ruderman} Clifford Cheung, Lawrence J. Hall, David Pinner, Joshua T. Ruderman, JHEP {\bf 1305}(2013)100.
\bibitem{nicolas} Nicolas Bernal, JCAP {\bf 0908}(2009)022.

\bibitem{bernal}Nicolas Bernal, Abdelhak Djouadi, Pietro Slavich, JHEP0707:016(2007).


\bibitem{giudice2} Gian F. Giudice, Alessandro Strumia, Nucl.Phys. B858 (2012) 63-83.



\bibitem{martin} Stephen P. Martin, Michael T. Vaughn, Phys.Lett. B318 (1993) 331-337.


\bibitem{nugaugino1}J. R. Ellis, K. Enqvist, D. V. Nanopoulos, K. Tamvakis, Phys. Lett. B155 (1985) 381.

\bibitem{nugaugino2} M. Drees, Phys. Lett. B158 (1985) 409.

\bibitem{twoloop1} M. E. Machacek and M. T. Vaughn, Nucl. Phys. B 222, 83 (1983);
                   Nucl. Phys. B 236, 221 (1984); Nucl. Phys. B 249, 70 (1985).

\bibitem{twoloop2} S. P. Martin and M. T. Vaughn, Phys. Rev. D 50, 2282 (1994).



\bibitem{protondecay}  Makoto Miura [Super-Kamiokande Collab.],ICHEP2010.
                       Y. Suzuki et al. [TITAND Working Group Collaboration], hep-ex/0110005.
\bibitem{protondecay2} J. Hisano, D. Kobayashi, T. Kuwahara and N. Nagata, JHEP {\bf 1307}, 038 (2013) [arXiv:1304.3651 [hep-ph]];\\
                       J. Hisano, T. Kuwahara and N. Nagata, Phys.\ Lett.\ B {\bf 723}, 324 (2013) [arXiv:1304.0343 [hep-ph]].

\bibitem{protondecaylattice}
S.~Aoki {\it et al.}  [JLQCD Collaboration],
Phys.\ Rev.\ D {\bf 62}, 014506 (2000).

\bibitem{darksusy} P. Gondolo {\it et al.}, JCAP 07 (2004) 008.
The code is available from http://www.physto.se/~edsjo/darksusy.

\bibitem{FWY}Fei Wang, Wenyu Wang, Jin Min Yang, Eur. Phys. J. C46:521-526(2006).

\bibitem{obliquep} G. Altarelli and R. Barbieri, Phys. Lett. B 253, 161 (1991);
M. E. Peskin, T. Takeuchi, Phys. Rev. D 46, 381 (1992).
\bibitem{stuconstraints} LEP and SLD Collaborations, Phys. Rept. 427 (2006) 257.
\bibitem{stuformula} J. Cao and J. M. Yang, JHEP 0812, 006 (2008).

\bibitem{effpotential}  M. Binger, Phys. Rev. D73, 095001 (2006).

\bibitem{nmssmtools} U. Ellwanger et al., JHEP 0502, 066 (2005).





\bibitem{Djoudi}   A.~Djouadi and M.~Drees, \PLB484, 183 (2000);
    G.~Belanger {\it et al.}, Comput.\ Phys.\ Commun.\  {\bf 180}, 747 (2009).

\bibitem{Carena} M.~S.~Carena {\it et al.}, \NPB577, 88 (2000).

\bibitem{Hisano:2010ct}
  J.~Hisano, K.~Ishiwata and N.~Nagata, arXiv:1007.2601 [hep-ph].

\bibitem{lattice}   H.~Ohki {\it et al.}, \PRD78, 054502 (2008);
    D.~Toussaint and W.~Freeman, \PRL103, 122002 (2009);
    J.~Giedt, A.~W.~Thomas and R.~D.~Young, \PRL103, 201802 (2009).

\bibitem{lux} D.S. Akerib et al. [LUX Collaboration], arXiv:1310.8214 [astro-ph.CO].

\bibitem{PDG} J. Beringer et al. (Particle Data Group), Phys. Rev. D86, 010001 (2012)

\bibitem{MStopyukawa} Giuseppe Degrassi, Stefano Di Vita, Joan Elias-Miro, Jose R. Espinosa, Gian F. Giudice, Gino Isidori, Alessandro Strumia,
 JHEP1208(2012)098.




\bibitem{lodone} Pier Paolo Giardino, Paolo Lodone, Mod. Phys. Lett. A, Vol. 29, No. 19 (2014) 1450099.


\end{thebibliography}
\end{document}